%% file: ACHIversion.tex
\documentclass[sigconf]{acmart}
\AtBeginDocument{%
  }

\usepackage{xcolor}

\usepackage{graphicx}    % 插图
\usepackage{subcaption}  % 支持子图
\usepackage{booktabs}
\usepackage{longtable}
\usepackage{float} 
\usepackage{booktabs}
\usepackage{tabularx}
\usepackage{array}
\usepackage{caption}
\usepackage{placeins}

\setcopyright{acmlicensed}
\copyrightyear{2026}
\acmYear{2026}
\setcopyright{cc}
\setcctype{by}
\acmConference[CHI EA '26]{Extended Abstracts of the 2026 CHI Conference on Human Factors in Computing Systems}{April 13--17, 2026}{Barcelona, Spain}
\acmBooktitle{Extended Abstracts of the 2026 CHI Conference on Human Factors in Computing Systems (CHI EA '26), April 13--17, 2026, Barcelona, Spain}
\acmDOI{10.1145/3772363.3798689}
\acmISBN{979-8-4007-2281-3/2026/04}

\begin{document}

\title{Misty Forest VR: Turning Real ADHD Attention Patterns into Shared Momentum for Youth Collaboration}

\author{Yibo Meng}
\affiliation{%
  \institution{Tsinghua University}
  \city{Beijing}
  %\state{Washington}
  \country{China}
}
\email{mengyb22@tsinghua.org.cn}

\author{Bingyi Liu}
\affiliation{%
  \institution{University of Michigan, Ann Arbor}
  \city{Ann Arbor}
  \state{Michigan}
  \country{United State}
}
\email{bingyi@umich.edu}

\author{Ruiqi Chen}
\affiliation{%
  \institution{University of Washington}
  \city{Seattle}
  \state{Washington}
  \country{United States}
}
\email{ruiqich@uw.edu}

\author{Yan Guan}
\affiliation{%
  \institution{Tsinghua University}
  \city{Beijing}
  %\state{Washington}
  \country{China}
}
\email{guany@tsinghua.edu.cn}
\renewcommand{\shortauthors}{Meng et al.}

\begin{abstract}
    Attention Deficit Hyperactivity Disorder (ADHD) remains highly stigmatized in many cultural contexts, particularly in China, where ADHD-related behaviors are often moralized rather than understood as neurodevelopmental differences. As a result, challenges of self-perception, social misunderstanding, and collaboration between ADHD and non-ADHD individuals remain largely unaddressed. We present Misty Forest, a VR-based collaborative game that explores ADHD through asymmetric co-play. The system translates empirically grounded ADHD behavioral patterns—such as fluctuating attention and time blindness—into complementary roles that require mutual coordination between players. Rather than compensating for deficits, the design treats cognitive difference as a source of interdependence.
    In a controlled study with mixed ADHD–non-ADHD dyads, Misty Forest led to higher task completion, increased self-acceptance among ADHD participants, improved ADHD knowledge, and greater empathy among non-ADHD players. These findings suggest that neurodiversity-centered interactive design can foster understanding, reciprocity, and inclusive collaboration.
\end{abstract}



\maketitle

\section{INTRODUCTION}
\input{sections/INTRODUCTION}
\section{RELATED WORK}
\input{sections/RELATED_WORK}

\input{sections/PRELIMINARY_STUDY}
{\section{GAME DESIGN}}
\input{sections/GAME_DESIGN}
{\section{EVALUATION}}
\input{sections/EVALUATION}
{\section{FINDINGS}}
\input{sections/RESULTS}

\section{DISCUSSION}

\input{sections/Discussion}
\section{CONCLUSION}
\input{sections/Conclusion}

\section{Acknowledgment of AI Use}
LLMs were used solely for minor language editing. 
All research content, analysis, and conclusions are entirely the authors’ own.

\bibliographystyle{ACM-Reference-Format}
\bibliography{Aref}
\clearpage
\appendix
\input{sections/appendix}

\end{document}

%% file: sections/INTRODUCTION.tex
Collaborative activities depend on shared timing, mutual anticipation, and coordinated attention.
For individuals with attention deficit hyperactivity disorder (ADHD), fluctuations in attention, impulse control, and time perception can disrupt these shared rhythms, often leading to breakdowns in everyday collaboration\cite{gvirts2021interpersonal,weissenberger2021time}.
In China, where ADHD remains underdiagnosed and highly stigmatized, such disruptions are frequently misinterpreted as laziness or poor self-discipline rather than coordination challenges \cite{neurolaunch2024adhd,gueorguieva2015cultural}.
As a result, individuals with ADHD often experience self-doubt and social exclusion, while non-ADHD peers may perceive behaviors such as distractibility or emotional dysregulation as signs of unreliability or disengagement \cite{gao2025comparative}.

Most ADHD interventions including medication, behavioral therapy, and digital training tools have largely been developed around managing individual symptoms.
From a design perspective, however, this emphasis leaves open important questions about how individuals with ADHD come to understand and accept their own attentional differences, and how mutual understanding and collaboration can be supported between ADHD and non-ADHD peers.
These challenges are especially acute in collectivist cultural contexts such as China, where academic conformity and emotional restraint often magnify the social cost of neurodivergent behaviors \cite{gueorguieva2015cultural}.

Game-based and immersive systems have been proposed as alternatives by leveraging intrinsic motivation and embodied interaction to scaffold attention and emotional regulation \cite{fang2023vradhd}.
However, most ADHD-oriented games in the Chinese market focus on individual cognitive training—such as working memory, time management, or task planning—and largely ignore the collaborative and sociocultural dimensions of ADHD-related challenges \cite{fang2023vradhd}.
Similarly, many VR-based empathy systems rely on short-term perspective-taking or sensory simulation to foster understanding \cite{yan2023executive,tan2019virtual}.
While such approaches can raise awareness, they often frame neurodivergence as a deficit to be corrected or temporarily experienced, reinforcing a helper--helped binary and offering limited support for sustained collaboration.

Together, this body of work points to an underexplored design opportunity: to move beyond individual regulation and momentary empathy toward collaborative mechanisms that allow attentional variability to actively shape coordination.
We address this gap through \textit{Misty Forest}, a two-player collaborative VR game co-designed with Chinese adolescents with ADHD, clinicians, and educators.
In the game, ADHD and non-ADHD players navigate a metaphorical foggy forest that externalizes attentional fluctuation and timing mismatch.
Rather than equalizing performance, \textit{Misty Forest} employs asymmetric roles and rhythm-based handovers that require players to adapt to each other’s attentional states, reframing attentional variability as a negotiable collaborative rhythm rather than an individual deficit.

%% file: sections/RELATED_WORK.tex
\textbf{ADHD Support Beyond Individual Regulation}
HCI research has increasingly moved from deficit-oriented models of ADHD toward
neurodiversity- and context-aware approaches. Prior systems have introduced
attention-training games, conversational agents, and everyday scaffolds to help
individuals manage focus, emotional regulation, and task rhythms
\cite{penuelas2022video, bul2018serious, tan2019virtual, cuber2024examining}.
More recent work adopts strengths-based framing, embedding support within family,
classroom, or workplace contexts rather than treating ADHD as an isolated
impairment \cite{spiel2022adhd, stefanidi2023children, park2024collaborative, kim2022workplace}.
However, most of these systems remain intra-personal, focusing on individual
self-regulation with limited attention to collaborative interaction with
non-ADHD peers.

\textbf{Immersive Empathy and Experiential Simulation}
Immersive VR has been widely used to foster empathy and perspective-taking through
first-person viewpoints, sensory distortion, and narrative role-play
\cite{tan2019virtual, chen2023design,zhao2025immersive}. While such systems can raise awareness and
inclusive design intent \cite{kim2018empath, lowy2023toward, meng2026whalesong}, prior work cautions
that short-term or scripted simulations may oversimplify lived experience or
reinforce stereotypes without participatory grounding
\cite{bennett2019promise, tigwell2021nuanced}. This limitation is particularly
salient for ADHD, where attention fluctuates dynamically through cycles of
overload, focus bursts, and recovery \cite{yan2023executive}.

\textbf{Design Gap}
Across ADHD-support and empathy-oriented systems, a key gap remains: translating
ADHD-related behavioral variability into shared coordination structures rather
than treating it as an individual deficit or momentary experience. This gap is
further amplified in East Asian cultural contexts, where ADHD-related behaviors
are often moralized through norms of self-discipline and productivity
\cite{gueorguieva2015cultural}. \textit{Misty Forest} addresses this challenge by
embedding attentional rhythms directly into collaborative VR mechanics, enabling
ADHD and non-ADHD players to negotiate difference through play.

%% file: sections/GAME_DESIGN.tex
\subsection{System Overview and Narrative}
\begin{figure*}[t]
    \centering
    \begin{subfigure}{0.48\linewidth}
        \centering
        \includegraphics[width=\linewidth]{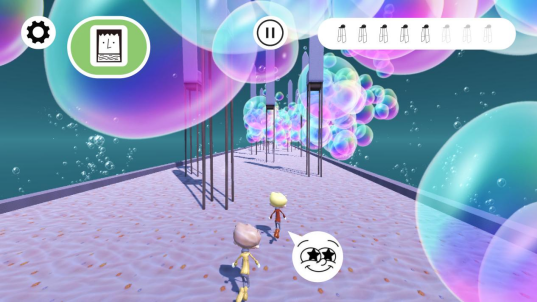}
        \caption{Supportive environmental cues.}
    \end{subfigure}
    \hfill
    \begin{subfigure}{0.48\linewidth}
        \centering
        \includegraphics[width=\linewidth]{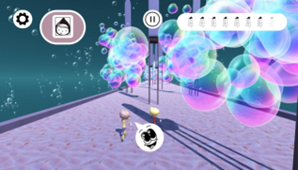}
        \caption{Distracting and unstable conditions.}
    \end{subfigure}
    \caption{The same gameplay context is experienced differently by ADHD and non-ADHD players, revealing coordination challenges that motivate asymmetric collaboration.}
    \label{fig:scene_comparison}
\end{figure*}

\textit{Misty Forest} is a two-player collaborative VR game designed to support coordination between individuals with ADHD and those without.
Adopting a neurodiversity perspective, the system treats attentional variability not as a deficit but as a resource for collaboration.
The game was co-designed with Chinese adolescents with ADHD, clinicians, and educators, shaped by formative interviews on stigma and misaligned expectations in everyday collaboration.

Players navigate a fog-filled forest that externalizes attentional fluctuation through unstable paths, salient distractions, and timing-sensitive obstacles.
A dyad consisting of one ADHD and one non-ADHD player must traverse the environment together, relying on communication and coordination to progress.
The narrative framing positions neither role as deficient; instead, differences in timing, planning, and perception are surfaced as shared coordination challenges that must be negotiated through play.

%\subsection{Key Mechanics: Rhythm-Based Asymmetric Collaboration}

%At the core of \textit{Misty Forest} is an asymmetric collaboration model grounded in attentional rhythm.
%Insights from early prototyping and interviews revealed recurring cycles of distraction, overload, and brief bursts of intense focus among players with ADHD.
%Rather than suppressing these fluctuations, the system translates them into complementary roles.

%The ADHD player assumes the role of the \textit{Exploder}, capable of activating short bursts of high-focus action to rapidly clear obstacles and advance the team.
%Following each burst, the player enters a brief pause.
%The non-ADHD player takes on the role of the \textit{Supporter}, providing strategic scaffolding by marking paths, stabilizing environmental conditions, and regulating pace.

%Gameplay follows a cyclical rhythm of \textit{support – burst – handover – pause}, requiring continuous communication and mutual adaptation.
%Support actions are intentionally triggered rather than automatic, emphasizing active empathy over passive assistance.
%Through this structure, attentional variability is reframed from an individual limitation into a shared coordination problem, where success depends on recognizing and working with cognitive difference in real time.

\subsection{Key Mechanics: Rhythm-Based Asymmetric Collaboration}

Our design is grounded in behavioral data collected from ADHD players in early VR prototypes. Logging and interviews revealed recurring attentional rhythms characterized by brief high-focus “bursts” followed by rapid depletion and increased error rates. Rather than suppressing this fluctuation, \textit{Misty Forest} operationalizes it as the structural basis of collaboration.

Each level consists of segmented parkour paths containing unstable platforms, branching intersections, dynamic obstacles, and salient distractions. Progression is gated: neither player can complete the level independently.

The ADHD player (\textit{Exploder}) can activate short high-speed burst phases that automatically clear dense obstacle clusters and open forward paths. Bursts are time-limited and followed by a mandatory cooldown period during which the player cannot advance.

The non-ADHD player (\textit{Supporter}) cannot clear major obstacles but can strategically scaffold the environment via ray-based interactions—marking optimal routes, reducing visual clutter, slowing obstacle timing, and stabilizing unstable platforms. Supports must be manually triggered and are limited in duration.

Interdependence is structurally enforced: certain obstacles require burst-clearing before becoming traversable, while others remain too unstable without prior support activation. Misaligned timing leads to stalled progression or cascading errors.

Environmental elements translate cognitive contrast into coordination demands: fog restricts visibility, unstable “stool” paths amplify motor instability, and salient distractors compete for attention. Successful play depends on synchronizing burst timing with strategic support, producing a recurring rhythm of \textit{support – burst – handover – pause}.

%% file: sections/EVALUATION.tex
We conducted a mixed-methods, controlled study to evaluate whether \textit{Misty Forest} (1) improves self-awareness and self-acceptance among adolescents with ADHD, (2) increases understanding and empathy among non-ADHD peers, and (3) supports usable cross-group collaboration through rhythm-based asymmetric roles.

\textbf{Design.} The study used a pre--post design with three conditions to isolate the role of collaboration structure:
\begin{itemize}
    \item \textbf{Experimental (Mixed-Pair Collaboration):} ADHD and non-ADHD participants formed mixed dyads and played the game collaboratively.
    \item \textbf{Control 1 (Same-Group Collaboration):} ADHD participants were paired with ADHD participants, and non-ADHD participants were paired with non-ADHD participants; all played collaboratively.
    \item \textbf{Control 2 (Individual Play):} Participants played the game individually without a teammate.
\end{itemize}
This design allowed us to compare mixed-pair collaboration against (i) collaboration without neurotype diversity and (ii) no collaboration.

\textbf{Participants.} We recruited $N=60$ participants (30 clinically diagnosed with ADHD; 30 without ADHD), aged 10--24. Participants were recruited via Chinese social platforms (e.g., WeChat, Xiaohongshu, Bilibili). All participants provided informed consent (guardian consent for minors), and the study was approved by the institutional ethics board.

\textbf{Procedure.} The study consisted of three stages:
\begin{itemize}
    \item \textbf{T1 (Pre-test):} Participants completed a baseline ADHD knowledge questionnaire \cite{sciutto2007evaluating}. ADHD participants additionally completed a self-perception / self-acceptance measure \cite{ryff1989happiness}. Non-ADHD participants completed a self-determination / tolerance measure (SDS) \cite{deci1985general}.
    \item \textbf{T2 (Gameplay):} Participants received a brief tutorial and then completed the gameplay session in their assigned condition (mixed pairs, same-group pairs, or individual play). 
    \item \textbf{T3 (Post-test \& Interview):} Participants repeated the ADHD knowledge questionnaire; ADHD participants repeated the self-acceptance measure and non-ADHD participants repeated SDS. Participants also completed post-session semi-structured interviews. The experimental group additionally completed a game experience questionnaire \cite{ijsselsteijn2013game}.
\end{itemize}

\textbf{Measures.} We examined four outcome categories:
(i) ADHD knowledge \cite{sciutto2007evaluating};
(ii) ADHD self-acceptance/self-awareness \cite{ryff1989happiness};
(iii) non-ADHD tolerance/empathy toward ADHD (SDS) \cite{deci1985general};
(iv) user experience/usability (GEQ) \cite{ijsselsteijn2013game}.
We also tracked task-level performance indicators such as completion outcomes.

\textbf{Analysis.} Quantitative results were summarized as pre--post changes by condition (mixed pairs vs.\ same-group pairs vs.\ individual play). %Interview transcripts were analyzed using thematic analysis \cite{braun2006using} to capture how asymmetric roles shaped coordination, perceived agency, and empathy during play.
Interview transcripts were analyzed using thematic analysis \cite{braun2006using}. Two researchers independently conducted open coding on an initial subset of transcripts to construct a preliminary codebook, which was iteratively refined across the full dataset until thematic saturation was reached. All transcripts were then dual-coded based on the finalized codebook, achieving 85\% intercoder agreement, with discrepancies resolved through discussion.

%% file: sections/RESULTS.tex
\textbf{Overall Study Outcomes.}
Mixed ADHD--non-ADHD collaboration consistently outperformed homogeneous and individual play, leading to higher task completion, increased self-acceptance among ADHD participants, greater tolerance and understanding among non-ADHD participants, and more stable collaborative performance. These outcomes were jointly supported by quantitative measures and participant accounts.

\textbf{Finding 1: Mixed collaboration enabled successful task completion and coordination.}
All mixed pairs in the experimental condition completed the game (10/10), compared to substantially lower completion rates in Control~1 (8/20) and Control~2 (1/20). Participants attributed success to recognizing and adapting to complementary cognitive rhythms, describing collaboration as shared coordination rather than individual assistance.

\textbf{Finding 2: Role-based agency increased self-acceptance among ADHD participants.}
ADHD participants in the experimental group showed a large increase in self-acceptance scores (from $M{=}7.1$ to $M{=}12.2$), significantly exceeding both control groups (time $\times$ group interaction: $F(2,54){=}26.64$, $p{<}.001$; see Fig.~2). 

Interview data suggested that the \textit{recharge} mechanic enabled players to reinterpret hyperfocus as contribution rather than deficit. As one participant noted, \textit{“It is not my problem; it is part of who I am.”} Others described feeling “accomplished” when using focused bursts to overcome obstacles, reflecting a shift toward capability-oriented self-perception.

\textbf{Finding 3: Mixed play fostered tolerance and understanding among non-ADHD participants.}
Non-ADHD participants in mixed pairs showed the largest gains in tolerance toward ADHD (SDS gain-score comparison: $F(2,27){=}17.68$, $p{<}.001$). 

Participants described an attribution shift away from moralized labels (e.g., ``lazy'') toward rhythm- and environment-sensitive interpretations of attention. One participant reflected, \textit{“I realized it’s not their fault that they cannot control it.”} Others reported recognizing ADHD-related traits as strengths within collaboration.

\textbf{Finding 4: Collaborative gameplay supported mutual learning about ADHD.}
Both ADHD and non-ADHD participants in the experimental group demonstrated significant gains in ADHD knowledge (ADHD: $t(9){=}-9.00$, $p{<}.001$; non-ADHD: $t(9){=}-8.51$, $p{<}.001$), exceeding learning outcomes in control conditions. Learning was described as emerging through embodied experience and negotiation rather than explicit instruction.

\textbf{Finding 5: High usability supported empathy while revealing design refinement opportunities.}
Participants reported high usability and engagement across gameplay dimensions. At the same time, they suggested that clearer mapping between level metaphors and ADHD traits could further strengthen the system’s educational clarity, indicating directions for future design refinement.
\begin{figure}[htbp]
    \centering
    \includegraphics[width=0.4\textwidth]{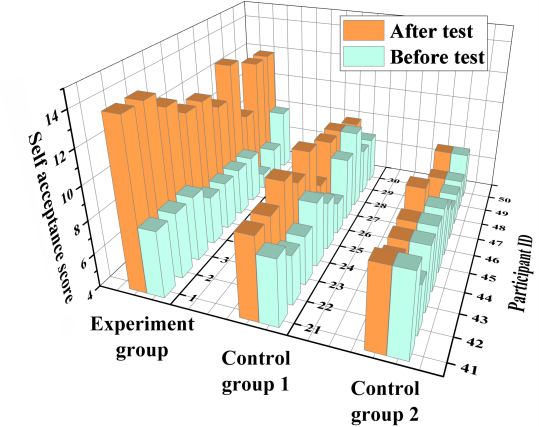}
    \hfill
    \includegraphics[width=0.35\textwidth]{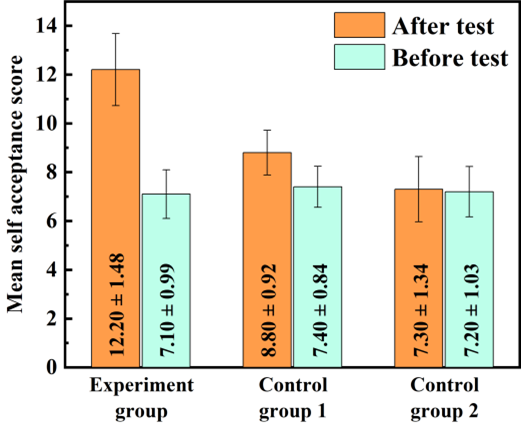}
    \caption{Self-acceptance scores of ADHD users in the experimental group, control group~1, and control group~2. Individual user (up); mean of each group (down).}
    \label{fig:self_acceptance_scores}
\end{figure}

%% file: sections/Discussion.tex
\textbf{From Deficit Compensation to Reciprocal Collaboration}
Prior HCI work has often framed ADHD as a deficit addressed through behavioral training or
functional compensation~\cite{fang2023vradhd, zhang2024study, oh2024diagnosis}. While effective for task
support, such approaches frequently position neurodivergent users as recipients rather than
co-contributors~\cite{spiel2022adhd, lowy2023toward}.

\textit{Misty Forest} instead embeds attentional variability into collaborative mechanics.
Asymmetric roles allow ADHD players to contribute through time-critical bursts of focus, while
non-ADHD players support pacing and coordination. Rather than minimizing difference, this design
treats cognitive diversity as complementary.

Across quantitative and qualitative results, mixed ADHD--non-ADHD pairs showed higher task
completion, stronger role reciprocity, and increased agency among ADHD players. These findings
suggest that designing for reciprocal asymmetry can move collaborative systems beyond accommodation
toward interdependence.

\textbf{Embodied and Culturally Situated Empathy} 
Empathy has long been emphasized as a foundation for inclusive systems, yet many approaches rely on narrative explanation or observational perspective-taking~\cite{kim2018empath, bennett2019promise}. Such strategies may support cognitive understanding but often fall short of fostering affective resonance or behavioral change~\cite{corrigan2023immersive, oh2022effectiveness}.

\textit{Misty Forest} instead fosters embodied empathy through immersive role reversal and task-bound interaction. Non-ADHD players, when inhabiting ADHD roles, experienced fluctuating attention, sensory overload, and disrupted temporal rhythms, prompting a shift from moralized interpretations of behavior toward rhythm- and environment-sensitive understanding. Crucially, empathy was not isolated from action: players were required to collaborate under asymmetric constraints, translating experiential understanding into supportive behavior during gameplay.

These effects were amplified through culturally situated metaphors, such as classroom-inspired disciplinary scenes familiar within Chinese educational contexts, where ADHD-related behaviors are often moralized~\cite{stefanidi2023children, silva2023unpacking}. By grounding empathy in shared sociocultural experiences, the system enabled participants to reflect on how educational norms and disciplinary practices shape perceptions of cognitive difference. This suggests that empathy-oriented collaborative systems may benefit from being both embodied and culturally contextualized, linking affective experience with situated social meaning.

\textbf{Limitations and Future Work}
This study has several limitations. Participants were drawn from a specific cultural context, and collaboration took place within structured, goal-oriented tasks that may not fully reflect the ambiguity of everyday social interaction. While the system emphasized empowerment, some participants experienced multiplayer collaboration as socially demanding, pointing to tensions between inclusion and emotional safety. Future work could explore adaptive role complexity, broader cultural settings, and alternative collaboration formats to better accommodate diverse comfort levels.

%% file: sections/Conclusion.tex
Misty Forest is a culturally grounded VR system that uses asymmetric collaboration to support self-acceptance among individuals with ADHD and foster empathy with non-ADHD peers. By structuring teamwork around complementary cognitive rhythms, the system reframes neurodivergent traits as collaborative strengths. Mixed-methods findings suggest that activating difference rather than compensating for deficits can promote affective understanding and socially meaningful collaboration, especially within stigmatized cultural contexts.

%% file: sections/appendix.tex
\subsection*{Appendix 1: Information about the Participants in the Evaluation}

\begin{table}[!h]
\centering
\caption{Information about the participants in the evaluation (1--30).}
\label{tab:participants-1}
\setlength{\tabcolsep}{2.5pt}
\renewcommand{\arraystretch}{1.1}
\begin{tabularx}{\columnwidth}{c c c c c >{\raggedright\arraybackslash}X}
\toprule
ID & ADHD (Y/N) & Gender & Urban/Rural & Age & Highest Education \\
\midrule
1  & Y & F & U & 19 & High School \\
2  & Y & F & U & 15 & Primary School \\
3  & Y & M & U & 10 & Primary School \\
4  & Y & F & R & 12 & Primary School \\
5  & Y & M & U & 22 & Bachelor's Degree \\
6  & Y & M & R & 24 & Bachelor's Degree \\
7  & Y & F & R & 23 & Bachelor's Degree \\
8  & Y & F & U & 17 & High School \\
9  & Y & F & R & 17 & Junior High School \\
10 & Y & M & R & 19 & High School \\
11 & N & M & U & 22 & High School \\
12 & N & F & R & 22 & High School \\
13 & N & F & R & 21 & High School \\
14 & N & M & R & 19 & High School \\
15 & N & M & U & 24 & Bachelor's Degree \\
16 & N & F & U & 23 & High School \\
17 & N & F & R & 24 & Bachelor's Degree \\
18 & N & F & R & 22 & Bachelor's Degree \\
19 & N & M & U & 22 & Bachelor's Degree \\
20 & N & F & R & 17 & Junior High School \\
21 & Y & M & U & 12 & Primary School \\
22 & Y & F & R & 18 & Junior High School \\
23 & Y & M & U & 23 & High School \\
24 & Y & F & R & 23 & High School \\
25 & Y & M & U & 23 & High School \\
26 & Y & F & R & 24 & Bachelor's Degree \\
27 & Y & F & U & 24 & High School \\
28 & Y & M & R & 19 & Junior High School \\
29 & Y & M & R & 19 & High School \\
30 & Y & F & R & 17 & High School \\
\bottomrule
\end{tabularx}
\end{table}
\begin{table}[!h]
\centering
\caption{Information about the participants in the evaluation (31--60).}
\label{tab:participants-2}
\setlength{\tabcolsep}{2.5pt}
\renewcommand{\arraystretch}{1.1}
\begin{tabularx}{\columnwidth}{c c c c c >{\raggedright\arraybackslash}X}
\toprule
ID & ADHD (Y/N) & Gender & Urban/Rural & Age & Highest Education \\
\midrule
31 & N & F & U & 16 & High School \\
32 & N & F & U & 23 & High School \\
33 & N & M & U & 24 & High School \\
34 & N & M & R & 24 & Bachelor's Degree \\
35 & N & F & U & 23 & High School \\
36 & N & M & R & 23 & Junior High School \\
37 & N & F & U & 17 & High School \\
38 & N & M & R & 18 & Bachelor's Degree \\
39 & N & F & U & 22 & High School \\
40 & N & M & U & 23 & Junior High School \\
41 & Y & F & R & 24 & High School \\
42 & Y & F & R & 23 & Junior High School \\
43 & Y & M & U & 24 & High School \\
44 & Y & M & U & 24 & High School \\
45 & Y & M & R & 19 & High School \\
46 & Y & F & R & 17 & Bachelor's Degree \\
47 & Y & F & U & 17 & High School \\
48 & Y & F & R & 19 & High School \\
49 & Y & M & U & 14 & Primary School \\
50 & Y & M & R & 17 & High School \\
51 & N & F & U & 22 & Primary School \\
52 & N & M & R & 21 & High School \\
53 & N & F & U & 13 & Primary School \\
54 & N & M & R & 22 & Bachelor's Degree \\
55 & N & F & U & 19 & High School \\
56 & N & M & R & 21 & Junior High School \\
57 & N & M & U & 22 & High School \\
58 & N & F & U & 22 & High School \\
59 & N & F & U & 23 & High School \\
60 & N & F & R & 17 & High School \\
\bottomrule
\end{tabularx}
\end{table}

\FloatBarrier
\subsection*{Appendix 2: ADHD Knowledge Questionnaire}

\begingroup
\setlength{\itemsep}{0.3em}
\setlength{\parskip}{0pt}

\begin{enumerate}

\item ADHD is a behavioral problem caused purely by laziness or a lack of willpower.\\
\textbf{Options:} True / False \quad
\textbf{Correct Answer:} False \quad
\textbf{Scoring Rule:} 1 point for answering ``False''

\item Individuals with ADHD cannot concentrate on anything.\\
\textbf{Options:} True / False \quad
\textbf{Correct Answer:} False \quad
\textbf{Scoring Rule:} 1 point for answering ``False''

\item The core symptoms of ADHD include inattention, hyperactivity, and impulsivity.\\
\textbf{Options:} True / False \quad
\textbf{Correct Answer:} True \quad
\textbf{Scoring Rule:} 1 point for answering ``True''

\item Many individuals with ADHD experience ``time blindness,'' meaning they have difficulty perceiving time accurately.\\
\textbf{Options:} True / False \quad
\textbf{Correct Answer:} True \quad
\textbf{Scoring Rule:} 1 point for answering ``True''

\item Environmental distractions (e.g., noise, visual stimuli) typically affect individuals with ADHD more than those without ADHD.\\
\textbf{Options:} True / False \quad
\textbf{Correct Answer:} True \quad
\textbf{Scoring Rule:} 1 point for answering ``True''

\item Individuals with ADHD cannot perform well under pressure.\\
\textbf{Options:} True / False \quad
\textbf{Correct Answer:} False \quad
\textbf{Scoring Rule:} 1 point for answering ``False''

\item The common ``procrastination'' in individuals with ADHD is entirely an attitude problem.\\
\textbf{Options:} True / False \quad
\textbf{Correct Answer:} False \quad
\textbf{Scoring Rule:} 1 point for answering ``False''

\item Frequently shaking a chair in class must be an ADHD student intentionally causing trouble.\\
\textbf{Options:} True / False \quad
\textbf{Correct Answer:} False \quad
\textbf{Scoring Rule:} 1 point for answering ``False''

\item Individuals with ADHD can sometimes enter a highly focused state called ``hyperfocus.''\\
\textbf{Options:} True / False \quad
\textbf{Correct Answer:} True \quad
\textbf{Scoring Rule:} 1 point for answering ``True''

\item Social misconceptions and stigma can exacerbate the psychological distress of individuals with ADHD.\\
\textbf{Options:} True / False \quad
\textbf{Correct Answer:} True \quad
\textbf{Scoring Rule:} 1 point for answering ``True''

\end{enumerate}
\endgroup

\subsection*{Appendix 3: Self-Acceptance Scale for People with ADHD}

\begin{table}[H]
\centering
\caption{Self-Acceptance Scale for People with ADHD}
\label{tab:self-acceptance}
\setlength{\tabcolsep}{3pt}
\renewcommand{\arraystretch}{1.1}
\begin{tabularx}{\columnwidth}{l >{\raggedright\arraybackslash}X}
\toprule
\textbf{Item} & \textbf{Statement (5-point Likert scale)} \\
\midrule
\multicolumn{2}{l}{\textit{Part A: Symptom Recognition \& Emotional Management}} \\
1 & I have difficulty maintaining focus on tasks for long periods and am easily distracted. \\
2 & I often make careless mistakes or overlook details. \\
3 & I often feel restless and have difficulty sitting still quietly. \\
4 & I often interrupt others or finish their sentences before they are done speaking. \\
5 & I have difficulty organizing tasks and managing time effectively. \\
6 & My emotions fluctuate greatly, and I easily become irritable or frustrated. \\
7 & I frequently procrastinate and find it hard to start tasks. \\
8 & I often forget daily routines, such as appointments or task deadlines. \\
\midrule
\multicolumn{2}{l}{\textit{Part B: Self-Acceptance \& Self-Esteem}} \\
9  & I am able to accept my characteristics related to attention and behavior. (Reverse Scored) \\
10 & I view these characteristics as a part of who I am, not as flaws. (Reverse Scored) \\
11 & I often dislike myself for not being able to focus as well as others. \\
12 & I feel proud of the creativity and energy I can exhibit in certain situations. (Reverse Scored) \\
13 & I worry that others will look down on me because of these characteristics. \\
14 & I believe in my ability to find ways to cope with life's challenges. (Reverse Scored) \\
15 & Overall, I am satisfied with myself. (Reverse Scored) \\
\bottomrule
\end{tabularx}
\end{table}